\documentclass[aps,amssymb,floatfix,prd,amsmath,preprintnumbers]{revtex4}
\usepackage{amsmath}
\usepackage{graphicx}  % Include figure files
\usepackage[font=scriptsize]{caption}
\usepackage{float}
\usepackage{dcolumn}   % Align table columns on decimal point
\usepackage{bm}% bold math
\usepackage{wasysym}
\begin{document}
%\input epsf.tex
%%%%%%%%%%%%
%%%%%%%%%%%

\title{\bf  Constraining Nuclear Symmetry Energy parameters from Neutron skin thickness of $^{48}$Ca}

\author{ S. K. Tripathy\footnote{Department of Physics, Indira Gandhi Institute of Technology, Sarang, Dhenkanal, Odisha-759146, India, E-mail: tripathy\_sunil@rediffmail.com}, D. Behera\footnote{1. Department of Physics, Indira Gandhi Institute of Technology, Sarang, Dhenkanal, Odisha-759146, India\\2. School of Physics, Sambalpur University, Jyotivihar, Sambalpur, Odisha-768019, India,\\E-mail:dipadolly@rediffmail.com},  T. R. Routray \footnote{Retired Professor, School of Physics, Sambalpur University, Jyotivihar, Sambalpur, Odisha-768019, India, E-mail:trr1@rediffmail.com}and B. Behera\footnote{Retired Professor, School of Physics, Sambalpur University, Jyotivihar, Sambalpur, Odisha-768019, India} 
}
\affiliation{ }

\begin{abstract}
In the present work, we use a finite range effective interaction to calculate the neutron skin thickness in $^{48}$Ca and correlate these quantities with the parameters of nuclear symmetry energy. Available experimental data on the neutron skin thickness in $^{48}$Ca are used to deduce information on the density slope parameter and the curvature symmetry parameter of the nuclear symmetry energy at saturation and at subsaturation densities. We obtained the constraints such as $54.5\leq L(\rho_0) \leq 97.5$ MeV and $47.3\leq L(\rho_c) \leq 57.1$ MeV for the density slope parameter. The constraints on the curvature symmetry energy parameter are obtained as $-170.7\leq K_{sym}(\rho_0) \leq -43.4$ MeV and $-80.8\leq K_{sym}(\rho_c) \leq 23.8$ MeV. A linear relation between the neutron skin thickness in $^{48}$Ca and in $^{2088}$Pb is obtained.
\end{abstract}
\maketitle
\textbf{PACS number}: 21.65.Ef,24.30.Cz
%\textbf{Keywords}:  
\section{Introduction} 

The nuclear symmetry energy (NSE), $E_s(\rho)$ is a fundamental quantity in the understanding of the  equation of state (EoS) of isospin asymmetric nuclear matter (ANM). The density dependence of NSE plays an important role in nuclear physics and astrophysics\cite{Lattimer2000, Steiner2005}. Since $E_s(\rho)$ is not a directly measurable quantity, there have been attempts from both the theoretical and experimental perspectives to understand the density dependence aspect of NSE \cite{Tsang2009, Tsang2012,Piek2012,Tamii2011, Maza2015, Zhang2014}. In fact, density dependence of $E_s(\rho)$ is the most uncertain part in the EoS \cite{Li2019a, Brown2000, Li2008} and mostly relies upon the determination of $E_s(\rho_0)$, its slope parameter $L(\rho_0)$ at saturation density $\rho_0$ and the curvature parameter $K_{sym}(\rho_0)$. While we have a fair knowledge on the value of $E_s(\rho_0)$  and its slope parameter $L(\rho_0)$, our present knowledge of $K_{sym}(\rho_0)$ is rather poor.  From different nuclear experiments and astrophysical observations prior to 2013, we have $E_s(\rho_0)=31.6\pm 2.7$ MeV and $L(\rho_0)=58.9\pm 16$ MeV \cite{Li2013}. The values  of $E_s(\rho_0)=31.7\pm 3.2$ MeV and $L(\rho_0)=58.7\pm 28.1$ MeV were obtained in a recent work \cite{Oertel2017}. Very recently, from a Bayesian analysis, the values  of $E_s(\rho_0)=31.7\pm 1.1$ MeV and $L(\rho_0)=59.8\pm 4.1$ MeV were obtained \cite{Drisch2020} 

Properties of finite nuclei provide stringent constraints on $E_s(\rho)$ and $L(\rho)$ at subsaturation densities. Fuchs and Wolter  from an analysis of different microscopic and phenomenological models, have obtained the NSE at a subsaturation density around $\rho_c\simeq 0.6\rho_0$ to be $E_s(\rho_c) \simeq 24$ MeV \cite{Fuchs2006}. From the properties of doubly magic nuclei, Brown tried to constrain the EoS at a density of $\rho_c=0.1$ fm$^{-3}$ \cite{Brown2013}. Zhang and Chen have obtained a tighter constraint on the symmetry energy at subsaturation density $\rho_c=0.11$ fm$^{-3}$ i.e. $E_s(\rho_c)=26.65\pm 0.20$ MeV \cite{Zhang2013} from an analysis of the binding energy difference of heavy isotope pairs. It is note here that the central density of heavy nucleus is around $0.11$ fm$^{-3}$ and a knowledge of the density slope parameter and the curvature symmetry parameter at this density is important in determining the density dependence of the NSE at low density region. 

The nuclear symmetry energy plays an important role in the formation of neutron skins in neutron-rich nuclei. The neutron skin thickness (NST), $\varDelta r_{np}=\left\langle r^2\right\rangle_n^{1/2}-\left\langle r^2\right\rangle_p^{1/2}$,  is used as a sensitive probe of NSE to improve our knowledge in the isovector channels of nuclear effective interaction at least in the subsaturation density region \cite{Trzcinska2001, Brown2007, Klos2007,Zenihiro2010}. In recent times, a lot of efforts have been made to correlate the NST with the parameters of nuclear symmetry energy. In fact, the NST is observed to have a linear relationship with the density slope parameter $L(\rho_0)$ \cite{Maza2011, Maza2018}. There have been a lot of efforts made to obtain the neutron skin thickness in $^{208}Pb$ and to constrain the density dependence of $E_s(\rho)$ from the results \cite{Klos2007,Zenihiro2010,Warda2009, Vinas2014, Maza2011, Mondal2016, Centelles2010}. The first run of the Lead Radius Experiment (PREX) measured the neutron skin thickness in $^{208}$Pb to be $\varDelta r_{np}=0.33^{+0.16}_{-0.18}~fm$ \cite{prex2012}. The PREX results have large error bars but the proposed PREX II is expected to reduce the error by a factor of 3 \cite{prex2}. The measurements from coherent pion photoproduction (the Mainz experiment) provide $\varDelta r_{np}(^{208}$Pb)=$0.15\pm 0.03(stat)^{+0.01}_{-0.03}(syst)~fm$ \cite{Tarbert2014}. The Calcium Radius Experiment (CREX) has also been approved and is ongoing at the Jefferson Lab. It is expected that, the CREX may reduce the error to $0.02$ fm \cite{prex2}. Using the coupled-cluster calculation, Hagen et al. obtained the neutron skin thickness  $\varDelta r_{np}$ in $^{48}$Ca as $0.12-0.15$ fm \cite{Hagen2015}. Also, from this calculation they have constrained the density slope parameter as $37.8\leq L(\rho_0)\leq 47.7$ MeV \cite{Hagen2015}. Birkhan et al. have determined the electric dipole strength in $^{48}$Ca from proton inelastic scattering experiments at RCNP, Osaka \cite{Birkhan2017}. From this experiments, they inferred the neutron skin thickness in $^{48}$Ca to be $\varDelta r_{np}(^{48}$Ca)=$0.17\pm 0.03$ fm. Very recently, Tanaka et al. \cite{Tanaka2020} obtained the neutron skin thickness in the isotopes of Ca from the interaction cross sections for $^{42-51}$Ca. The NST for $^{48}$Ca from these observations yielded $\varDelta r_{np}(^{48}$Ca)=$0.146\pm 0.048$ fm \cite{Tanaka2020}. Tagami et al. recently used Gogny-D1S Hatree-Fock-Bogoliubov model with angular momentum projection to calculate the neutron skin thickness in some Calcium isotopes and obtained $\varDelta r_{np}(^{48}$Ca)=$0.159- 0.190$ fm  \cite{Tagami2020}.  Xu et al. have carried out a Bayesian analysis on the measured and some speculated values of the neutron skin thickness in Sn isotopes, $^{208}$Pb and $^{48}$Ca to constrain the density dependence of nuclear symmetry energy \cite{Xu2020}. Amidst all these efforts to constrain the density slope parameter and the curvature symmetry parameter, our knowledge on the density dependence of NSE is still very poor  at subsaturation density region.

In the present paper, we calculate the neutron skin thickness of the doubly magic nuclei $^{48}$Ca using the EoSs constructed from a finite range effective interaction. Recent experimental constraints on the NST in $^{48}$Ca are used to constrain the parameters of nuclear symmetry energy at the saturation density and at a subsaturation density. The paper is organised as follows: in Section II, the basic formalism of the nuclear equation of state as obtained from the finite range effective interaction is presented. The method of constraining the interaction parameters is discussed in brief. In Section III, we calculate the neutron skin thickness of $^{48}$Ca using the finite range effective interactions within the frame work of droplet model. The correlation of the neutron skin thickness with the parameters of nuclear symmetry energy have been carried out. Experimental constraints on the neutron skin thickness in $^{48}$Ca are used to constrain the nuclear symmetry energy parameters. A linear relationship is obtained between the NST of $^{48}$Ca and $^{208}$Pb. The conclusion and summary of the present work are presented in Section-V.

\section{Basic Formalism}
\subsection{Finite range effective interaction and Nuclear Symmetry Energy}

We consider a finite range simple effective interaction (SEI)\cite{Behera2020}
\begin{equation}\label{eq:1}
v_{eff}(\textbf {r})=t_0(1+x_0P_{\sigma})\delta(\textbf {r})+\frac{1}{6}t_3(1+x_3P_{\sigma})\left[\frac{\rho(\textbf {R})}{1+b\rho(\textbf {R})}\right]^{\gamma}\delta (\textbf {r})+\left(W+BP_{\sigma}-HP_{\tau}-MP_{\sigma}P_{\tau}\right)f(r),
\end{equation}
where $f(r)$ is the form factor which may have either a Gaussian or Yukawa or an exponetial form. Here we consider a Yukawa form factor $f(r)=\frac{e^{-r/{\alpha}}}{r/{\alpha}}$, $\alpha$ being the range of the interaction. $\textbf{r}$ and $\textbf{R}$ are respectively the relative and centre of mass coordinates of the two interacting nucleons. $W, B, H$ and $M$ are the strength parameters of the Wigner, Bartlett, Heisenberg and Majorana components.  $P_{\sigma}$ and $P_{\tau}$ are the spin and isospin exchange operators respectively. The parameter $b$ takes care of the supara lumious behaviour at high density and  $\gamma$ determines the stiffness of the nuclear equation of state in symmetric nuclear matter (SNM). Other parameters of the interactions, $t_0,x_0,t_3,x_3$,  are adjusted so as to reproduce the saturation properties of SNM.  This SEI has already been used to study the momentum and density dependence of the isoscalar part of the nuclear mean field at zero and finite temperature \cite{Behera1998, Behera2002, Routray2000}, isovector part of the nuclear mean field at zero temperature \cite{Behera2005, Behera2007}, temperature dependence of nuclear symmetry energy \cite{Behera2009, Behera2011} and to calculate the half-lives of spherical proton emitters \cite{Routray2011}. The SEI with a Gaussian form factor for the finite range part of the effective interaction has been used in recent times  to address the problem of binding energy and charge radii of spherical nuclei \cite{Behera2013}, spin polarized neutron matter \cite{Behera2015}, deformation properties of nuclei \cite{Behera2016} and neutron star properties \cite{Routray2016,Pattnaik2018}.

The energy density $H(\rho,y_p, T)$ in ANM at a density $\rho$, proton fraction $y_p$ and temperature $T$ can be obtained from SEI as
\begin{eqnarray}
H(\rho,y_p, T) &=& \int \left[f_T^n(\textbf{k})+f_T^p(\textbf{k})\right]\left(c^2\hbar^2k^2+M^2c^4\right)~d^3k\nonumber\\
               &+& \frac{1}{2}\left[\frac{\varepsilon_0^l}{\rho_0}+\frac{\varepsilon_{\gamma}^l}{\rho_0^{\gamma+1}}\left(\frac{\rho}{1+b\rho}\right)^{\gamma}\right]\left(\rho_n^2+\rho_p^2\right)+\left[\frac{\varepsilon_0^{ul}}{\rho_0}+\frac{\varepsilon_{\gamma}^{ul}}{\rho_0^{\gamma+1}}\left(\frac{\rho}{1+b\rho}\right)^{\gamma}\right]\rho_n\rho_p\nonumber\\
               &+& \frac{\varepsilon_{ex}^{l}}{2\rho_0} \int\int \left[f_T^n(\textbf{k})f_T^n(\textbf{k}^{\prime})+f_T^p(\textbf{k})f_T^p(\textbf{k}^{\prime})g_{ex}(|\bf{k}-\bf{k}^{\prime}|)\right]~d^3k~d^3k^{\prime}\nonumber\\
               &+& \frac{\varepsilon_{ex}^{ul}}{2\rho_0} \int\int \left[f_T^n(\textbf{k})f_T^p(\textbf{k}^{\prime})+f_T^p(\textbf{k})f_T^n(\textbf{k}^{\prime})g_{ex}(|\bf{k}-\bf{k}^{\prime}|)\right]~d^3k~d^3k^{\prime}\label{eq:2},
\end{eqnarray}
where $f_T^{\tau}(\textbf{k}), \tau=n,p$ are the respective Fermi-Dirac distribution functions, $\Lambda=\frac{1}{\alpha}$ and $g_{ex}(|{\bf{k}}-{\bf{k}}^{\prime}|)=\frac{1}{1+\frac{|{\bf{k}}-{\bf{k}}^{\prime}|^2}{\Lambda^2}}$. The new parameters $\varepsilon^l_0$, $\varepsilon^{ul}_0$, $\varepsilon^{l}_{\gamma}$, $\varepsilon^{ul}_{\gamma}$, $\varepsilon^{l}_{ex}$ and $\varepsilon^{ul}_{ex}$ are related to the interaction parameters as
\begin{eqnarray}
\varepsilon^{l}_{0} &=& \rho_0\left[\frac{t_0}{2}(1-x_0)+4\pi \alpha^3\left(W+\frac{B}{2}-H-\frac{M}{2}\right)\right],\label{eq:3}\\
\varepsilon^{ul}_{0} &=& \rho_0\left[\frac{t_0}{2}(2+x_0)+4\pi \alpha^3\left(W+\frac{B}{2}\right)\right],\label{eq:4}
\end{eqnarray}
\begin{eqnarray}
\varepsilon^{l}_{\gamma} &=& \rho_0^{\gamma+1}\left[\frac{t_3}{12}(1-x_3)\right],\label{eq:5}\\
\varepsilon^{ul}_{\gamma} &=& \rho_0^{\gamma+1}\left[\frac{t_3}{12}(2+x_3)\right],\label{eq:6}\\
\varepsilon^{l}_{ex} &=& 4\pi \alpha^3\rho_0\left(M-\frac{W}{2}-B+\frac{H}{2}\right),\label{eq:7}\\
\varepsilon^{ul}_{\gamma} &=& 4\pi \alpha^3\rho_0\left(M+\frac{H}{2}\right).\label{eq:8}
\end{eqnarray}

The energy per particle in SNM is obtained at zero temperature ($T=0$) as
\begin{equation}
e_0(\rho)=\frac{3Mc^2}{8x_f^3}\left[2x_fu_f^3-x_fu_f-ln\left(x_f+u_f\right)\right]+\frac{\varepsilon_0}{2}\frac{\rho}{\rho_0}+\frac{\varepsilon_{\gamma}}{2}\frac{\rho}{\rho_0^{\gamma+1}}\left(\frac{\rho}{1+b\rho}\right)^{\gamma}+\frac{\varepsilon_{ex}}{2\rho_0}\rho J_0(\rho)\label{eq:9},
\end{equation}
where $x_f=\frac{\hbar k_f}{Mc}$, $u_f=\left(1+x_f\right)^{\frac{1}{2}}$. The Fermi momentum in SNM is given by  $k_f=\left(1.5\pi^2\rho\right)^{\frac{1}{3}}$. The functional $J_0(\rho)$ is given by
\begin{equation}
J_0(\rho) = \frac{\int\left(\frac{3j_1(k_fr)}{k_fr}\right)^2\frac{e^{-r/{\alpha}}}{r/{\alpha}}d^3r}{\int \frac{e^{-r/{\alpha}}}{r/{\alpha}}d^3r}\label{eq:10},
\end{equation}
where  $j_1(k_fr)$ is the first order spherical Bessel function and $\varepsilon_0=\frac{1}{2}\left(\varepsilon_{0}^{l}+ \varepsilon_{0}^{ul}\right), \varepsilon_{\gamma}=\frac{1}{2}\left(\varepsilon_{\gamma}^{l}+ \varepsilon_{\gamma}^{ul}\right)$, $\varepsilon_{ex}=\frac{1}{2}\left(\varepsilon_{ex}^{l}+ \varepsilon_{ex}^{ul}\right)$.

The zero temperature EoS in pure neutron matter (PNM) is obtained as
\begin{equation}
e_n(\rho)=\frac{3Mc^2}{8x_n^3}\left[2x_nu_n^3-x_nu_f-ln\left(x_n+u_n\right)\right]+\frac{\varepsilon_0^l}{2}\frac{\rho}{\rho_0}+\frac{\varepsilon_{\gamma}^l}{2}\frac{\rho}{\rho_0^{\gamma+1}}\left(\frac{\rho}{1+b\rho}\right)^{\gamma}+\frac{\varepsilon_{ex}^l}{2\rho_0}\rho J_{n}(\rho)\label{eq:11},
\end{equation}
where $x_n=\frac{\hbar k_n}{Mc}$, $u_n=\left(1+x_n\right)^{\frac{1}{2}}$. $k_n=\left(3\pi^2\rho\right)^{\frac{1}{3}}$ denotes the Fermi momentum in PNM. The functional $J_{n}(\rho)$ is expressed as
\begin{equation}
J_{n}(\rho) = \frac{\int\left(\frac{3j_1(k_nr)}{k_nr}\right)^2\frac{e^{-r/{\alpha}}}{r/{\alpha}}d^3r}{\int \frac{e^{-r/{\alpha}}}{r/{\alpha}}d^3r}\label{eq:12},
\end{equation}
where  $j_1(k_nr)$ is the first order spherical Bessel function. 

The nuclear symmetry energy, $E_s(\rho)$, is defined as

\begin{equation}
E_s(\rho)=\frac{1}{2!}\frac{\partial^2e(\rho,\delta)}{\partial\delta^2}\mid_{\delta=0}, \label{eq:13}
\end{equation}
and can also be expressed as the difference in the energy per particle in pure neutron matter $e_n(\rho)= e(\rho,\delta=1)$ and that in SNM,
\begin{equation}
E_s(\rho)=e_n(\rho)-e_0(\rho)\label{eq:14},
\end{equation}
where the contribution from higher order terms in $\delta$ is assumed to be small. With this definition of NSE, we can have from Eqs.\eqref{eq:9} and \eqref{eq:11}

\begin{eqnarray}
E_s(\rho)&=& \frac{3Mc^2}{8}\left[ \frac{2x_nu_n^3-x_nu_f-ln\left(x_n+u_n\right)}{x_n^3}-\frac{2x_fu_f^3-x_fu_f-ln\left(x_f+u_f\right)}{x_f^3}\right]\nonumber\\
&+& \frac{(\varepsilon_0^l-\varepsilon_0)}{2}\frac{\rho}{\rho_0}+\frac{(\varepsilon_{\gamma}^l-\varepsilon_{\gamma})}{2}\frac{\rho}{\rho_0^{\gamma+1}}\left(\frac{\rho}{1+b\rho}\right)^{\gamma}+\frac{[\varepsilon_{ex}^lJ_{n}(\rho)-\varepsilon_{ex}J_{0}(\rho)]}{2}\frac{\rho}{\rho_0}.\label{eq:15}
\end{eqnarray}

An expansion of NSE around saturation density $\rho_0$ reads as

\begin{eqnarray}
E_s(\rho) &=& E_s(\rho_0)+L\left(\frac{\rho-\rho_0}{3\rho_0}\right)
         +\frac{K_{sym}}{2}\left(\frac{\rho-\rho_0}{3\rho_0}\right)^2+\cdots \label{eq:16},
\end{eqnarray}
where $L(\rho_0)=3\rho_0\frac{\partial E_s(\rho)}{\partial \rho}|_{\rho=\rho_0}$ and $K_{sym}=9\rho_0^2\frac{\partial^2 E_s(\rho)}{\partial \rho^2}|_{\rho=\rho_0}$ are respectively the slope and curvature parameters of $E_s(\rho)$ at $\rho_0$. It is obvious from the above expansion that the density dependence of the nuclear symmetry energy relies upon the exact determination of the parameters $L(\rho_0)$ and $K_{sym}(\rho_0)$. 

We may expand the NSE around a subsaturation density $\rho_c<\rho_0$ as 
\begin{equation}
E_s(\rho)\approx E_s(\rho_c)+L(\rho_c)\varepsilon+\frac{K_{sym}(\rho_c)}{2!}\varepsilon^2+\mathcal{O}(\varepsilon^3)\label{eq:17},
\end{equation}
where $\varepsilon=\frac{\rho-\rho_c}{3\rho_c}$. $L(\rho_c)=3\rho_c\frac{dE_s(\rho)}{d\rho}|_{\rho=\rho_c}$ is the density slope parameter and $K_{sym}(\rho_c)=9\rho_c^2\frac{d^2E_s(\rho)}{d\rho^2}|_{\rho=\rho_c}$ is the curvature parameter at the reference density $\rho_c$.

\subsection{Fixation of interaction parameters}
The interaction parameters of SEI are adjusted so as to obtain viable equations of state for the SNM and PNM and to have a good description of the momentum dependence of nuclear mean field. The complete description of SNM requires only the knowledge of six parameters $\gamma, b, \alpha, \varepsilon_0, \varepsilon_{\gamma}$ and $\varepsilon_{ex}$.  However, the equation of state in PNM requires the splitting of the strength parameters $\varepsilon_0, \varepsilon_{\gamma}$ and $\varepsilon_{ex}$ into like ($l$) and unlike ($ul$) channels. We do not have any available experimental or empirical constraints for this splitting. Behera et al. have  constrained the parameter $\varepsilon_{ex}^l$ as $\varepsilon_{ex}^l=\frac{2}{3}\varepsilon_{ex}$ \cite{Behera2009} which allows the neutron effective mass in neutron-rich matter to pass over the proton effective mass. Once the splitting of $\varepsilon_{ex}$ into $\varepsilon_{ex}^l$ and $\varepsilon_{ex}^{ul}$ is fixed, we require the nuclear symmetry energy  $E_s(\rho_0)$ and its slope $E_s^{\prime}(\rho_0)=\rho\frac{dE_s(\rho)}{d\rho}|_{\rho=\rho_0}=\frac{1}{3}L(\rho_0)$ at saturation density to obtain the splitting of the other two strength parameters $\varepsilon_{0}$ and $\varepsilon_{\gamma}$ into like and unlike components. The details of constraining the parameters required for SNM are given in Refs. \cite{Behera2009, Behera2020} where the standard values $Mc^2=939~ MeV$, energy per nucleon in SNM $e_0(\rho_0)=923~ MeV$, $\left(c^2\hbar^2k_{f_0}^2+M^2c^4\right)^{\frac{1}{2}}=976~ MeV$ corresponding to the saturation density $\rho_0=0.1658~fm^{-3}$ are used. In order to constrain the parameters required for PNM, we follow the procedure as described in Refs. \cite{Behera2020, Behera2020a}.  The SEI predicts an incompressibility in normal nuclear matter, $K=240~MeV$ corresponding to $\gamma=0.5$ and an effective mass in SNM as $\frac{m^*}{M}=0.67$. The nuclear symmetry energy from the constructed EoSs provide a good description of its density dependence for a wide range of nuclear matter density. The NSE at a sub-saturation density $\rho_c\simeq \frac{2}{3}\rho_0\approx 0.11$ fm$^{-3}$ for all the sets of interaction parameters is obtained to be $E_s(\rho_c)=26.65$ MeV. At a density around twice the normal nuclear matter density, $E_s(2\rho_0)$ lies close to the limit $E_s(2\rho_0)= 46.9 \pm 10.1~MeV $, a constraint obtained from the analysis of astrophysical observations for a constant maximum mass of $M_{max}=2.01 M_{\astrosun}$ and radius $R_{1,4}=12.83$ \cite{Zhang2019, Li2019} of massive neutron stars GW170817.

\section{Neutron Skin Thickness in $^{48}Ca$}

In this section, we calculate the neutron skin thickness in $^{48}$Ca using the EoSs constructed from the finite range effective interaction (SEI) within the framework of droplet model Myers and Swiatecki \cite{Myers1980}. It is worth to mention here that, Myers and Swiatecki in their work \cite{Myers1980} have argued that, the droplet model results of the neutron skin thickness are almost equal to the results obtained by Hatree-Fock (HF) calculations. The reason behind the striking similarity between the DM and HF results lie in the fact that, the shell effects appearing in HF calculations may not be important for the discussion of neutron skin thickness \cite{Myers1980}. In a recent work, we have also calculated the neutron skin thickness of some nuclei using the finite range effective interactions in the framework of droplet model and obtained similar results to that of the HF calculations \cite{Behera2020a}. 

The neutron skin thickness of nuclei has been identified as a strong isovector indicator \cite{Reinhard2010}. In general, NST is defined as the difference between the rms radii for the density distribution of the neutrons and protons in the nucleus. Basing upon different contributions to NST, we can write
\begin{equation}
\varDelta r_{np}= \sqrt{\frac{3}{5}}\left[t-\frac{e^2Z}{70E_s(\rho_0)}+\frac{5}{2R}\left(b_n^2-b_p^2\right)\right], \label{eq:18}
\end{equation}
where $t$ is the distance between the neutron and proton radii of uniform sharp distributions, $b_n$ and $b_p$ are the surface widths of the neutron and proton profiles. Neglecting the shell correction within the purview of the droplet model, we can have

\begin{equation}
t=\frac{3}{2}r_0\frac{E_s(\rho_0)}{Q}\left(\frac{I-I_c}{1+x_A}\right)\label{eq:19}.
\end{equation}
where $r_0=\left(\frac{4}{3}\pi\rho_0\right)^{-1/3}$, $I=\frac{N-Z}{A}$ is the neutron-proton asymmetry in the nucleus and $I_c=\frac{e^2Z}{20E_s(\rho_0)R}$ is the Coulomb correction to the symmetry energy coefficient. The factor $x_A=\frac{9}{4}\frac{E_s(\rho_0)}{Q}A^{-1/3}$ is associated with the ratio of the surface symmetry energy to the volume symmetry energy of semi infinite nuclear matter. $Q=\frac{9}{4}\left(\frac{E_s(\rho_0)}{a_{sym}(A)}-1\right)^{-1}E_s(\rho_0)A^{-1/3}$ is the surface stiffness parameter that measures the resistance of the nucleus against separation of neutrons from protons to form a skin. Assuming the validity of  $a_{sym}(A)=E_s(\rho_A)$ and using the expansion $E_s(\rho_A) \simeq E_s(\rho_0)+L(\rho_0)\epsilon_A+\frac{K_{sym}(\rho_0)}{2}\epsilon_A^2 $, Eq.\eqref{eq:19} can be reduced to \cite{Behera2020}
\begin{equation}
t\simeq -2r_0\epsilon_A\beta\left(1+\frac{K_{sym}(\rho_0)}{2L(\rho_0)}\epsilon_A\right)A^{1/3}(I-I_c)\label{eq:20},
\end{equation}
where $\beta=\frac{L(\rho_0)/3}{E_s(\rho_0)}=\frac{E_s^{\prime}(\rho_0)}{E_s(\rho_0)}$ and $\epsilon_A=\frac{\rho_A-\rho_0}{3\rho_0}$. There appears to be a clear linear correlation between between the bulk part of the NST in finite nuclei and some isovector indicators such as $1-\frac{a_{sym}(A)}{E_s(\rho_0)}, \beta $ and $\frac{K_{sym}(\rho_0)}{E_s(\rho_0)}$.  In a recent work, we have used 16 sets of interaction parameters by varying the $E_s(\rho_0)$ and $L(\rho_0)$ so as to reproduce the symmetry energy at the central density of $^{208}$Pb as $26.65$ MeV. In that work, we have found that, EoSs with same value of NSE at saturation density $\rho_0$ may have different slopes. In view of this, the ratio $\beta=\frac{L(\rho_0)}{3E_s(\rho_0)}$ has a critical role in deciding the quantity $t$ and consequently $\varDelta r_{np}$ rather than $L(\rho_0)$. The surface contribution to the neutron skin thickness $\varDelta r_{np}^{surf}$ can be evaluated from the neutron and proton density profiles. Many authors have considered that $b_n\simeq b_p$, so that $\varDelta r_{np}^{surf} \simeq 0$. However, Warda et al. \cite{Warda2009} have obtained a linear relation $\varDelta r_{np}^{surf}=\left(0.3\frac{E_s(\rho_)}{Q}+0.07\right)I$ fm for the surface contribution to the neutron skin thickness. With the inclusion of the surface contribution as prescribed by Warda et al.\cite{Warda2009}, the NST can now be expressed as

\begin{equation}
\varDelta r_{np}= \sqrt{\frac{3}{5}}\left[-2r_0\epsilon_A\beta\left(1+\frac{K_{sym}(\rho_0)}{2L(\rho_0)}\epsilon_A\right)A^{1/3}(I-I_c)-\frac{e^2Z}{70E_s(\rho_0)}+\left(\sqrt{\frac{3}{5}}\frac{E_s(\rho_0)}{2Q}+0.0904\right)I\right].\label{eq:21}
\end{equation}
\begin{figure}
\includegraphics[width=0.7\textwidth]{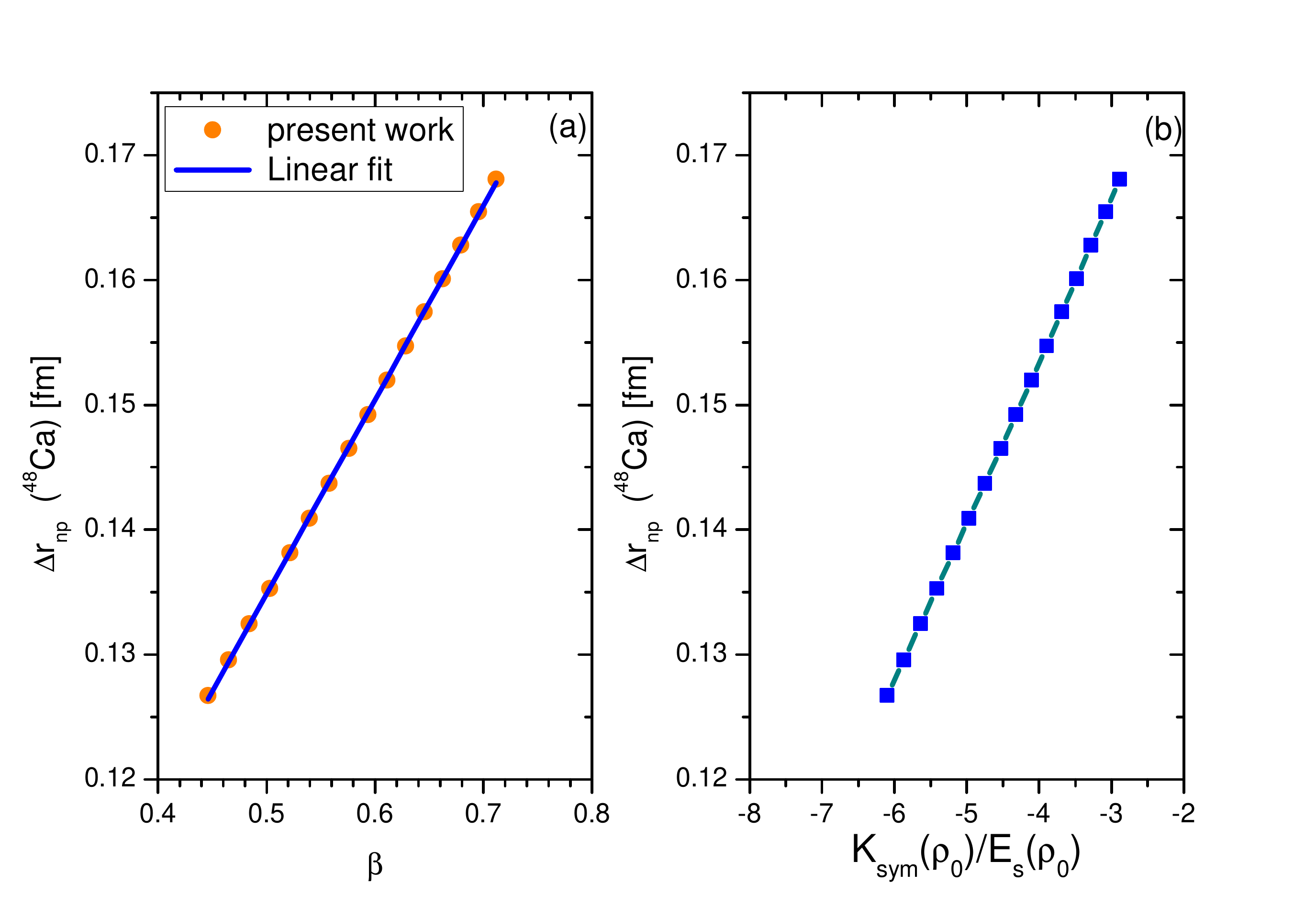}
\caption{ (a) The neutron skin thickness of $^{48}$Ca is shown as a function of $\beta$. A linear fit to the values is also shown in the figure. (b) The neutron skin thickness of $^{48}$Ca is shown as a function of $\frac{K_{sym}(\rho_0)}{E_s(\rho_0)}$.}
\end{figure}

It is obvious from the above expression \eqref{eq:21} that, $\varDelta r_{np}$ has a linear relationship with $\beta$ and $\frac{K_{sym}(\rho_0)}{E_s(\rho_0)}$. In Figures 1(a) and 1(b), we plot the neutron skin thickness of $^{48}$Ca calculated using the SEI in the framework of droplet model as function of $\beta$ and  $\frac{K_{sym}(\rho_0)}{E_s(\rho_0)}$. Linear plots are obtained for these correlations.  A linear fit provides us the relations

\begin{eqnarray}
\varDelta r_{np}(^{48}Ca)&=& 0.057+0.156~\beta~~ \text{fm},\label{eq:22}\\
\varDelta r_{np}(^{48}Ca)&=& 0.212+0.0004~K_{sym}(\rho_0)~~ \text{fm}.\label{eq:23}
\end{eqnarray}

A high resolution measurement of the electric dipole polarisability $\alpha_D$ in $^{48}$Ca at RCNP, Osaka predicted the neutron skin thickness in $^{48}$Ca as $\varDelta r_{np}(^{48}$Ca)=$0.17\pm 0.03$ fm \cite{Birkhan2017}. Very recently, by measuring the interaction cross section for $^{48}$Ca scattering on a target at RIKEN, Tanaka et al. have obtained $\varDelta r_{np}(^{48}$Ca)=$0.146\pm 0.048$ fm \cite{Tanaka2020}. In a recent work \cite{Behera2020a}, we have used the experimental values of the neutron skin thickness in $^{208}$Pb to constrain the nuclear symmetry energy parameters. In the present work, we use similar methods to constrain the nuclear symmetry parameters from the experimental values of the $\varDelta r_{np}(^{48}$Ca).  In Figure 2(a), we plot the neutron skin thickness as calculated using the SEI as function of the density slope parameter $L(\rho_0)$ at saturation density. The experimentally extracted regions  from the Osaka-RCNP and the RIKEN measurements are also shown  in the figure for comparison. A comparison of our results with the Osaka-RCNP results constrains the slope parameter $L(\rho_0)$ in the range $54.5-102$ MeV. On the other hand, the RIKEN results, constrain the slope parameter in the range $21.3\leq L(\rho_0)\leq 97.5$ MeV. It is to note here that, the NST as calculated from all the sets of the SEI are encompassed by the experimental region of the Osaka-RCNP results but the RIKEN region is compatible with the sets of SEI with $E_s(\rho_0)\geq 34$ MeV.

In Figure 2(b), the neutron skin thickness in $^{48}$Ca calculated from SEI is shown as a function of the curvature parameter at saturation density $K_{sym}(\rho_0)$ and compared with the results from Osaka-RCNP and RIKEN measurements. While the Osaka-RCNP results constrain the curvature parameter in the range $-170.7\leq K_{sym}(\rho_0) \leq -29.2$ MeV, the RIKEN results constrain it in the range $-269.8\leq K_{sym}(\rho_0) \leq -43.4$ MeV.

\begin{figure}
\includegraphics[width=0.7\textwidth]{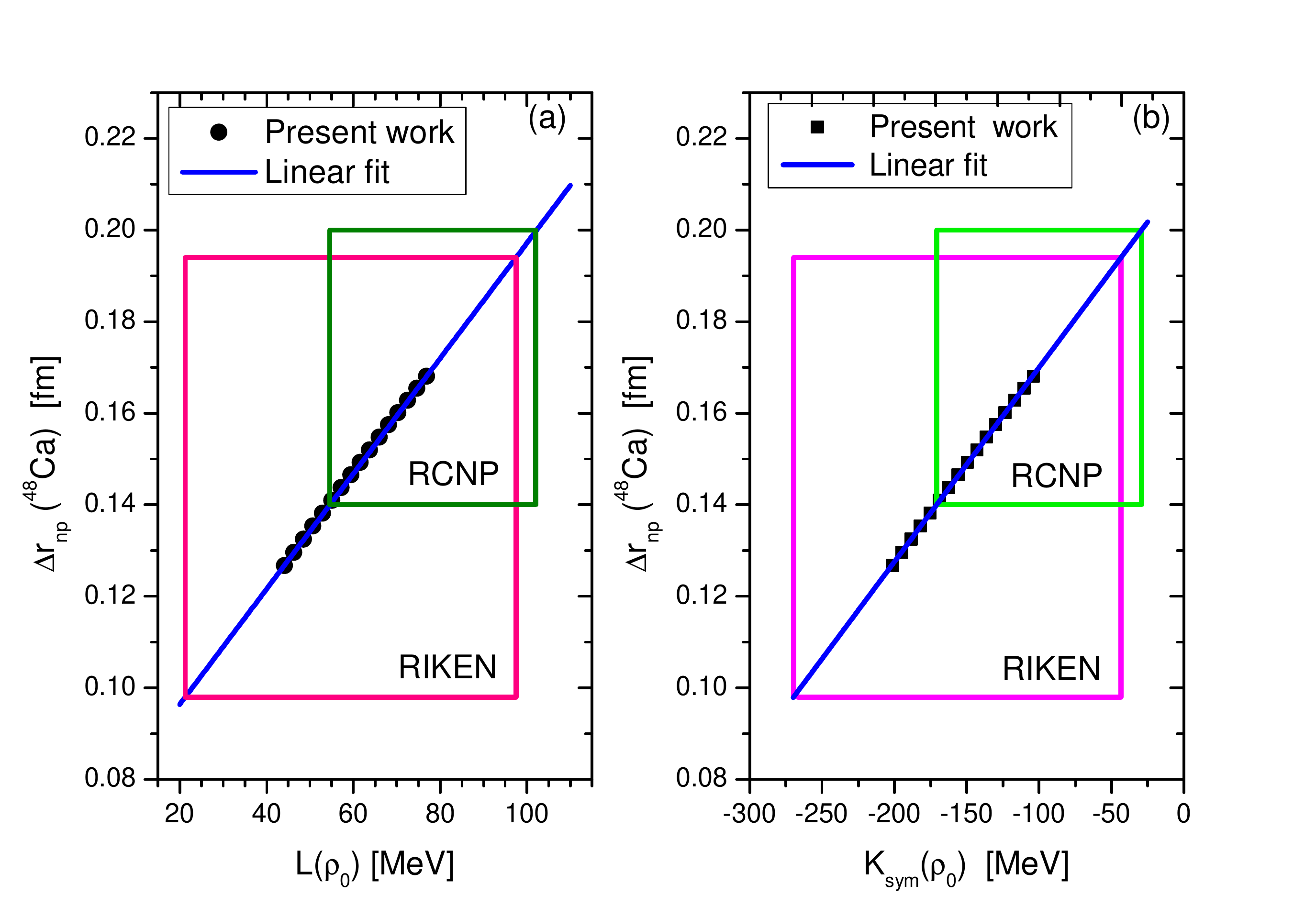}
\caption{ The neutron skin thickness in $^{48}$Ca is plotted as a function of (a) $L(\rho_0)$ and (b) $K_{sym}(\rho_{0})$. The experimental regions for $\varDelta r_{np}(^{48}$Ca) from the Osaka-RCNP measurements \cite{Birkhan2017} and the RIKEN measurements \cite{Tanaka2020} are shown for comparison.}
\end{figure}

\begin{figure}
\includegraphics[width=0.7\textwidth]{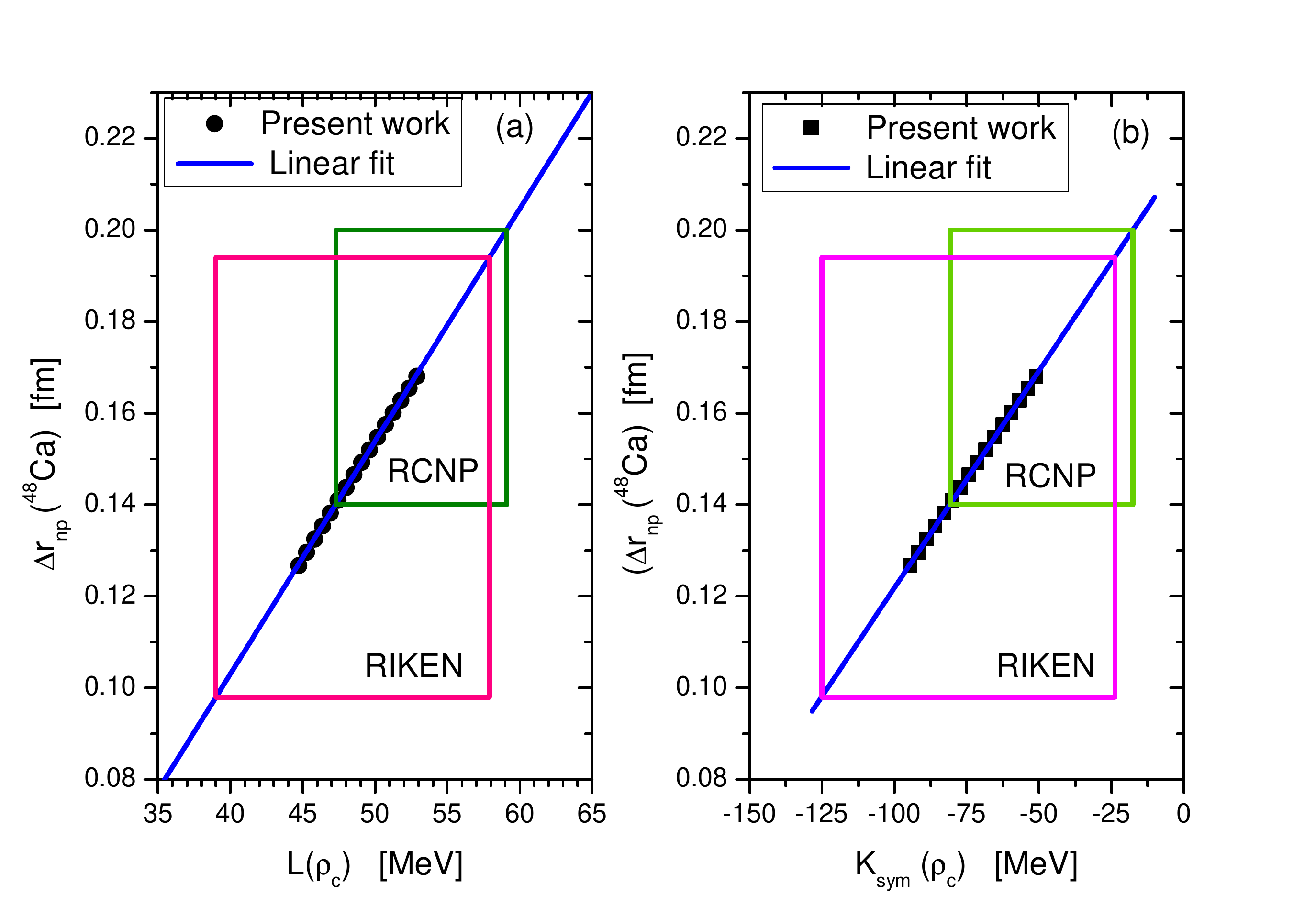}
\caption{ he neutron skin thickness in $^{48}$Ca is plotted as a function of (a) $L(\rho_c)$ and (b) $K_{sym}(\rho_{c})$. The experimental regions for $\varDelta r_{np}(^{48})$Ca from the Osaka-RCNP measurements \cite{Birkhan2017} and the RIKEN measurements \cite{Tanaka2020} are shown for comparison..}
\end{figure}

We wish to correlate the neutron skin thickness in $^{48}$Ca with the density slope parameter at a reference density $\rho_c<\rho_0$.  Replacing $\rho$  by $\rho_0$ in Eq. \eqref{eq:17} and keeping upto 2nd order in $\varepsilon$, we get \cite{Behera2020}

\begin{equation}
t=2r_0\varepsilon \beta^{\prime}\left[1+\frac{1}{2}\frac{K_{sym}(\rho_c)}{L(\rho_c)}\varepsilon\right]A^{1/3}(I-I_c), \label{eq:24}
\end{equation}
where $\beta^{\prime}=\frac{L(\rho_c)}{3E_s(\rho_0)}$  and  $\varepsilon=\frac{\rho_0-\rho_c}{3\rho_c}$. Consequently, the neutron skin thickness is expressed as \cite{Behera2020}
\begin{equation}
\varDelta r_{np}=\sqrt{\frac{3}{5}}\left[2r_0\varepsilon\beta^{\prime}\left(1+\frac{K_{sym}(\rho_c)}{2L(\rho_c)}\varepsilon\right)A^{1/3}(I-I_c)-\frac{e^2Z}{70E_s(\rho_0)}+\left(\sqrt{\frac{3}{5}}\frac{E_s(\rho_0)}{2Q}+0.0904\right)I\right].\label{eq:25}
\end{equation}

It is obvious from the above expression in Eq.\eqref{eq:24} that, the neutron skin thickness has a linear relationship with the parameters $\beta^{\prime}$ and $K_{sym}(\rho_c)$. From the calculations of $\varDelta r_{np}(^{48}$Ca) using the SEI, we may infer the linear relations as

\begin{eqnarray}
\varDelta r_{np}(^{48}Ca)&=& -0.104+0.489~\beta^{\prime}~~ \text{fm},\label{eq:26}\\
\varDelta r_{np}(^{48}Ca)&=& 0.215+0.0007~K_{sym}(\rho_c)~~ \text{fm}.\label{eq:27}
\end{eqnarray}

The Eq.\eqref{eq:26} can be easily translated as $\varDelta r_{np}(^{48}$Ca)= $-0.017+0.003~L(\rho_c)$ which provides a linear relation between the NST and the density slope parameter at a subsaturation density. In Figures 3(a) and (b), we show $\varDelta r_{np}(^{48}$Ca) as function of $L(\rho_c)$ and $K_{sym}(\rho_c)$ respectively. The results of Osaka-RCNP and RIKEN measurements are also shown in the figure for comparison. The Osaka-RCNP results constraints the density slope parameter at the subsaturation density in a tighter range as $47.3 \leq L(\rho_c) \leq 59.1$ MeV and the RIKEN results constrains the parameter as $39 \leq L(\rho_c) \leq 57.1$ MeV. We obtain the constraints on the parameter $K_{sym}(\rho_c)$ as $-80.8 \leq K_{sym}(\rho_c)\leq -17.6$ from a comparison with the results of Osaka-RCNP and as $-125 \leq K_{sym}(\rho_c)\leq -23.8$ from a comparison with the results of Osaka-RCNP. 

\begin{figure}
\includegraphics[width=0.7\textwidth]{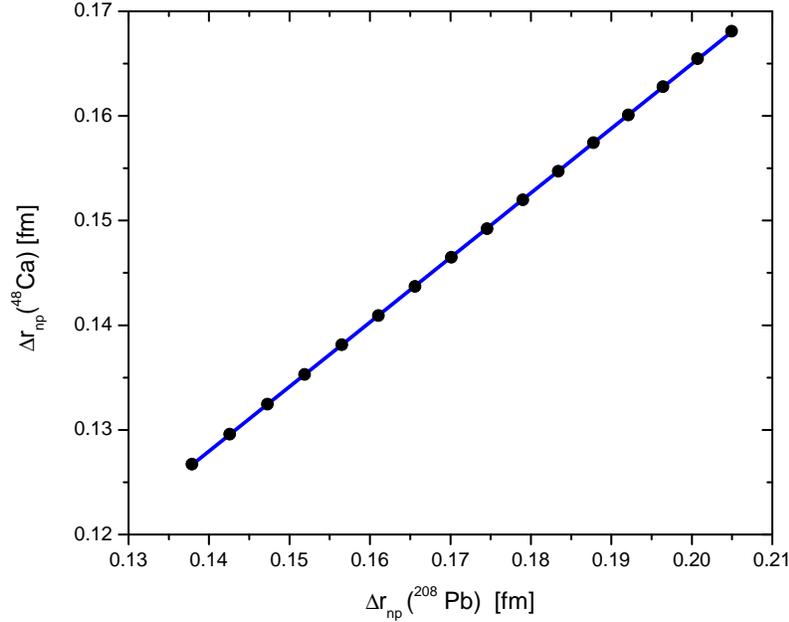}
\caption{ The neutron skin thickness in $^{48}$Ca is correlated with the that in $^{208}Pb$.}
\end{figure}

In Figure 4, we show the calculations of NST using the finite range effective interactions for $^{48}$Ca as function of the NST in $^{208}$Pb. An obvious linear correlation is obtained for these quantities in the figure. A linear fit to the results reads as
\begin{equation}
\varDelta r_{np}(^{48}Ca)= 0.0416+0.6169~\varDelta r_{np}(^{208}Pb).\label{eq:28}
\end{equation}
The central value of the NST in $^{208}$Pb as obtained in the first run of PREX \cite{prex2012} is $0.33$ fm. Using this value in Eq.\eqref{eq:28}, we may have a crude idea about the CREX result with the estimated error of CREX as
\begin{equation}
\varDelta r_{np}(^{48}Ca)= 0.245\pm 0.02~~\text{fm}.\label{eq:29}
\end{equation}
This result is large as compared to the experimental estimates from Osaka-RCNP and RIKEN. Experiments with hadronic probes constrained the NST in $^{208}$Pb as $\varDelta r_{np}= 0.16\pm (0.02)_{(\text{stat})}\pm (0.04)_{(\text{syst}}$ fm \cite{Klos2007} and  $\varDelta r_{np}= 0.211 ^{+0.054}_{-0.063}$ (Osaka-RCNP)\cite{Zenihiro2010} and measurements from coherent pion photoproduction yield  a value $\varDelta r_{np}(^{208}Pb)= 0.15\pm 0.03$ fm  (Mainz experiment)\cite{Tarbert2014}. A substitution of the central values of these experimental results yield $\varDelta r_{np}(^{48}$Ca)= $0.14\pm 0.02$ fm and $\varDelta r_{np}(^{48}$Ca)= $0.134\pm 0.02$ fm respectively which are compatible to the Osaka-RCNP and RIKEN results. 

\begin{figure}
\includegraphics[width=0.7\textwidth]{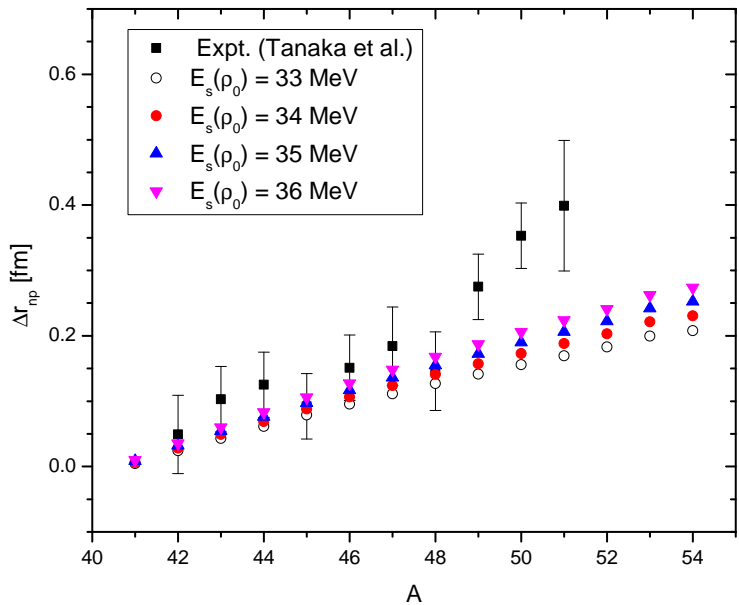}
\caption{ The neutron skin thickness in calcium isotopes plotted as function of mass number. The experimental results of Tanaka et al. \cite{Tanaka2020} for the NST in calcium isotopes are also  shown in the figure for comparison.}
\end{figure}

The experimental results of the neutron skin thickness in different calcium isotopes are available. Very recently, Tanaka et al. have determined the NST in $^{42-51}$Ca from the measurement of interaction cross section \cite{Tanaka2020}. We calculate the neutron skin thickness for some of the calcium isotopes by using the finite range effective interaction (SEI) and plot them as function of mass number in Figure 5 for four different sets of interaction parameters. The experimental results of Tanaka et al. are also shown in the figure for comparison. One may observe that, in general, the neutron skin thickness in calcium isotopes increases with an increase in the mass number. The theoretical calculations from SEI follow the experimental trend of Tanaka et al. and reproduce the results for the isotopes $^{42-48}$Ca. However, for the isotopes with mass number greater than 48, our results from SEI are underestimated as compared to the experimental values. It appears from the figure that, the sets of the finite range effective interaction  with higher values of $E_s(\rho_0)$ are more favoured for the calculations of neutron skin thickness in calcium isotopes.
\section{Summary and Conclusion}
In the present work, we have calculated the neutron skin thickness of $^{48}$Ca using some recently constructed EoSs from finite range effective interaction (SEI) in the framework of droplet model. The EoSs from SEI provide a good description of the nuclear symmetry energy at a subsaturation density ($\rho_c<\rho_0$), saturation density $\rho_0$ and at a suprasaturation density ($2\rho_0$) and therefore they are suitable for applications to a wider range of density. The finite range effective interactions predict the neutron skin thickness of $^{48}$Ca in the range $0.128-0.169$ fm with a spread of about $0.04$ fm. Experimental constraints on $\varDelta r_{np}(^{48}$Ca) are available from the Osaka-RCNP and the RIKEN measurements. We used these experimental constraints to constrain some of the nuclear symmetry energy parameters such as the density slope parameter and the curvature symmetry energy parameter at the saturation density and at the subsaturation density.  While the results of the Osaka-RCNP measurements constrain the slope parameter $L(\rho_0)$ in the range $54.5-102$ MeV, the RIKEN results constrain the slope parameter in the range $21.3\leq L(\rho_0)\leq 97.5$ MeV. The constraints as obtained for the curvature symmetry energy parameter from the Osaka-RCNP and the RIKEN results are  $-170.7\leq K_{sym}(\rho_0) \leq -29.2$ MeV and $-269.8\leq K_{sym}(\rho_0) \leq -43.4$ MeV respectively. At a subsaturation density $\rho_c\simeq 0.11$ fm$^{-3}$, the constraints as obtained from a comparison of the experimental results and the present calculations using the finite range effective interactions are $47.3 \leq L(\rho_c) \leq 59.1$ MeV (Osaka-RCNP) and $39 \leq L(\rho_c) \leq 57.1$ MeV (RIKEN). For the curvature symmetry energy parameter, we obtained $-80.8 \leq K_{sym}(\rho_c)\leq -17.6$ from a comparison with the results of Osaka-RCNP and  $-125 \leq K_{sym}(\rho_c)\leq -23.8$ from a comparison with the results of Osaka-RCNP. The NST as calculated from all the sets of the SEI are encompassed by the experimental region of the Osaka-RCNP results and the RIKEN region is compatible with the sets of SEI with $E_s(\rho_0)\geq 34$ MeV.

From the calculations of the NST in $^{208}$Pb with the finite range effective interactions (SEI) within droplet model, we obtained a linear relationship between $\varDelta r_{np}(^{48}$Ca) and $\varDelta r_{np}(^{208}$Pb). This relation can be used to predict the result of CREX as $\varDelta r_{np}(^{48}$Ca)=$0.245\pm 0.02$ fm. This result is somewhat larger as compared to the recent experimental estimates from Osaka-RCNP and RIKEN which may be due to the large central value of $\varDelta r_{np}(^{208}$Pb) obtained in PREX. However, accurate determination of the neutron skin thickness in $^{208}$Pb can predict the CREX results with some accuracy from the obtained relationship.

\end{document}